# Brain Tumor Radiogenomic Classification: A Deep Learning Approach

Amr Mohamed, Mahmoud Rabea, Aya Sameh, Ehab Kamal
Systems and Biomedical Department, Faculty of Engineering Cairo University

*Abstract-* The RSNA-MICCAI brain tumor radiogenomic classification challenge[1] aimed to predict MGMT biomarker[2] status in glioblastoma through binary classification on Multi parameter mpMRI scans: T1w, T1wCE, T2w and FLAIR. The dataset is splitted into three main cohorts: training set, validation set which were used during training , and the testing were only used during final evaluation. Images were either in a DICOM format[3] or in png format[4]. different architectures were used to investigate the problem including the 3D version of Vision Transformer (ViT3D)[5], ResNet50[6], Xception[7] and EfficientNet-b3[8]. AUC was used as the main evaluation metric and the results showed an advantage for both the ViT3D and the Xception models achieving **0.6015** and **0.61745** respectively on the testing set. compared to other results, our results proved to be valid given the complexity of the task. further improvements can be made through exploring different strategies, different architectures and more diverse datasets.

## Introduction

The RSNA-MICCAI brain tumor radiogenomic classification challenge[1] was held back in 2021 in order to predict the status of the MGMT biomarker through a binary classification across different mpMRI scan types. MGMT promoter methylation[2] is an important biomarker for glioblastoma[9], the most common and aggressive form of brain cancer in adults.By introducing a Radiogenomic based imaging method, the process of detecting the presence of brain tumor shall be less invasive which will eventually improve the survival and prospects of patients with brain cancer. Also, knowing the methylation status helps guide treatment decisions, as tumors with methylation are more responsive to certain therapies.

While the output of this task is a the probability of glioblastoma which represents the model's confidence of that the given tumor is glioblastoma, the input of the model consisted of the Brain mpMRI scan where a separate model was trained for each scan type [T1w , T1wCE , T2w and FLAIR] then all models were ensemble to get the final prediction.

## Related Work

Brain tumor detection from MRI scan has seen great advancements over the past few years, Abdusalomov et al. [10] proposed a deep learning approach using pre-trained YOLOv7 for object detection, Bi-directional Feature Pyramid Network (BiFPN) for feature extraction, and Channel and Spatial Attention module (CBAM) for improved attention mechanisms. Their model achieved a remarkable 99.5% accuracy, outperforming traditional methods; however, its focus on large tumors and dependence on specific MRI scanners and acquisition protocols necessitate further investigation. Saeedi et al. [11] presented two convolutional deep learning methods for

MRI-based brain tumor detection, incorporating data augmentation and pre-processing. Their 2D CNN and Convolutional Auto-Encoder achieved competitive accuracy, but generalization to small tumors and robustness across varied scanners warrant enhancement. Yogananda et al.[12] introduced a 3D-dense-UNet architecture for determining glioma MGMT promoter methylation status, showcasing a mean cross-validation accuracy of 94.73%, sensitivity of 96.31%, and specificity of 91.66%. However, ongoing code-related errors require resolution. Furthermore, a novel radiogenomic classification approach by S. A. Qureshi[13] demonstrated a two-stage prediction system, achieving impressive results (96.84% accuracy, 96.08% sensitivity, 97.44% specificity) in predicting MGMT promoter methylation status for glioblastoma patients using a multi-omics fused feature space with k-NN and SVM classifiers. While promising, addressing redundant features and ensuring model robustness are identified areas for improvement in these evolving methodologies.

## Dataset

*3.1 Dataset Description*

The dataset is defined by three main cohorts:
- 468 training subjects
- 117 validation subjects
- 87 testing subjects

across different cohorts, the subjects are labeled by a five-digit number "eg. 00002" and inside each subject's folder, there are different types of mpMRI scan in a DICOM format. the exact types of mpMRI scans are:
- T1-weighted pre-contrast (T1w)
- T1-weighted post-contrast (T1wCE)
- T2-weighted (T2w)
- Fluid Attenuated Inversion Recovery (FLAIR)

*3.2 Preprocessing*

As a preprocessing step, we converted the DICOM images into a 3D Numpy array representing the height, width and depth of the

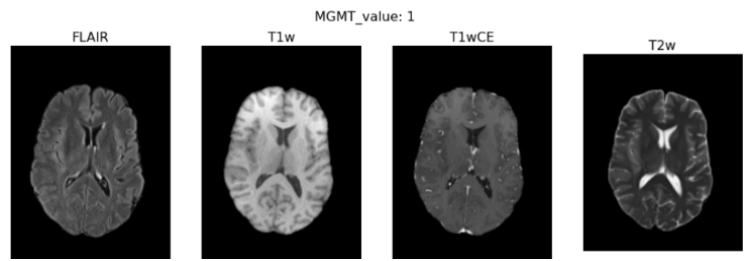

[1] Dataset Example

3D volume. Images were then resized to 256x256 with 64 frames for each image.

Value of Interest Lookup Table (VOI LUT) is applied to enhance the contrast of the DICOM images for better results. Finally, augmentation is applied for the training samples to increase the diversity of the dataset.
The following techniques were used for augmentation:
- 90 degrees clockwise rotation
- 90 degrees counterclockwise rotation
- 180 degrees rotation

For training EfficientNet-B3, ResNet50, and Xception model RSNA MICCAI PNG dataset on Kaggle was used[4]. This dataset contains the "DICOM data " of the RSNA-MICCAI Brain Tumor Radiogenomic Classification challenge in PNG format. Images were resized to 300 x 300 for training of EfficientNet-B3, as mentioned in Keras documentation[14]), and were resized to 512 x 512 for training of ResNet50[6] and Xception[7] models. Random data augmentation was applied during training of all three models to prevent overfitting:
- Horizontal flip
- Random rotation (± 36°)
- Random horizontal and vertical Translation (± 10% of the original dimension)

## Methodology

*4.1 3D ViT*
The ViT model was originally designed for handling two-dimensional data, as introduced in [5]. This model is then modified to handle three-dimensional input, i.e., each embedding is obtained by flattening a 3D patch rather than a 2D one.

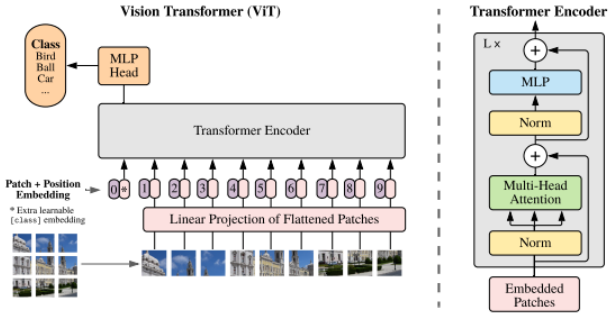

[2] original ViT

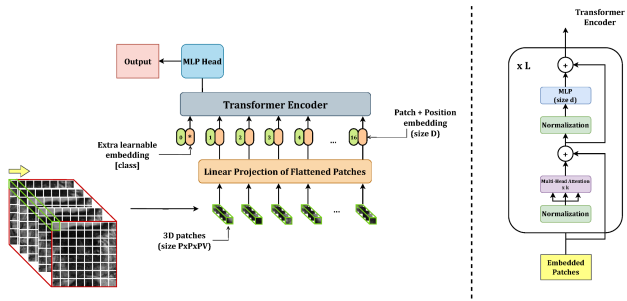

[3] ViT-3D

For our proposed solution, we trained 4 ViT-3D models on each mpMRI scan type [T1w, T1wCE, T2w, FLAIR] We compared two different architectures of ViT-3D. The first architecture uses an input of patches with dimension (16x16x16) and it Consists of two Transformer encoder blocks, each with 16 attention heads. and dropout rates of 0.1 are applied for both general and embedding layers. In contrast, the second architecture maintains the same hyperparameters (Two Transformer encoder blocks, 16 attention heads and dropout rates of 0.1) but adopts a larger patch size of (32x32x32). This adjustment of patch size was made to investigate the impact of increased spatial context on the model's ability to capture patterns within the mpMRI scans. Then we applied either simple averaging or stacking ensembles with logistic regression to obtain the final prediction probabilities.

*4.2  EfficientNet-B3, ResNet50, and Xception*

These complex and deep architectures were trained initially on large-scale labeled image dataset (ImageNet[15]) to learn standardized image properties. We used the pretrained models with fine tweaking to adapt to our task.

In this approach we trained a model for each type of mpMRI namely FLAIR, T1w, T1wCE, and T2w. Only the added layers were trained for 20 epochs while weights of the pretrained model were freezed. Then, we used a simple averaging method of the probability scores obtained from each model to get our final Prediction of whether the MGMT biomarker is present or not. We added the same layers for all three pretrained models.

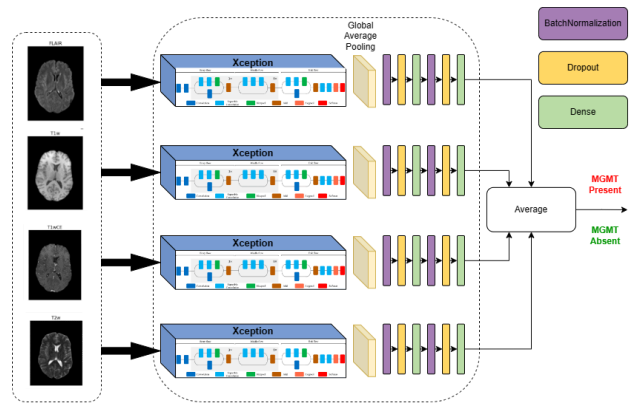

[4] Proposed architecture for detection of MGMT biomarker

**Results and Discussion**

We used the area under the curve (AUC) as our main metric to compare between the predicted probability and the observed target. The AUC is calculated on the validation dataset and the testing dataset respectively.

*5.1  3D ViT*

For different ensemble techniques, models were trained with different epoch numbers while maintaining the other parameters mentioned in the methodology  as it's. Early stopping was also applied and for each mpMRI scan type, models were saved according to the best loss given on the validation dataset.

The following Tables show the results of each technique emphasizing the best AUC score:

| Model | Ensemble Technique | Epoch Num | Validation AUC | Testing AUC |
| --- | --- | --- | --- | --- |
| ViT3D-16 | Simple Averaging | 10 | 0.6388 | 0.5837 |
| ViT3D-16 | Stacking | 10 | 0.5735 | 0.5782 |
| *ViT3D-32 | **Simple Averaging** | **10** | **0.5869** | **0.6015** |
| ViT3D-32 | Stacking | 10 | 0.5802 | 0.6009 |
| ViT3D-32 | Simple Averaging | 15 | 0.6314 | 0.5934 |

Table [1] Performance Metrics of the ViT-3D Models with 256 image size and 0.2 validation split

As observed, the best model was the one with simple averaging ensemble technique having slightly better results from the one with stacking. To further investigate the performance, two models were trained on a different validation split (0.1) and with different image sizes as shown in the following table:

| Model | image size | Epoch Num | Validation AUC | Testing AUC |
| --- | --- | --- | --- | --- |
| ViT3D-32 | 256 | 10 | 0.6521 | 0.6006 |
| ViT3D-32 | 512 | 10 | 0.7039 | 0.5971 |

Table [2] Performance Metrics with 0.1 validation split

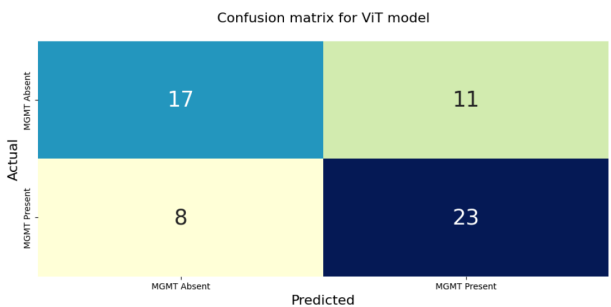

Figure [1] Confusion matrix for ViT3D model with image size 512

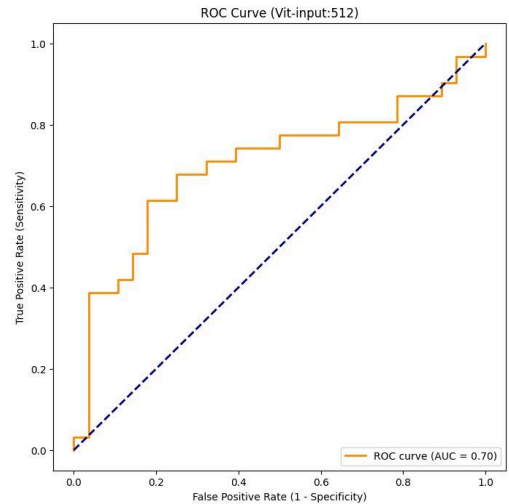

Figure [2] AUC curve for ViT3D model with image size 512

### 5.2 EfficientNet-B3, ResNet50, and Xception

In addition to the ViT model, we explored the performance of EfficientNet-B3, ResNet50, and Xception architectures in predicting MGMT promoter status. Early stopping and learning-rate decay were applied, and the models were saved based on the best loss observed on the validation dataset.

EfficientNet-B3 exhibited moderate performance, with a validation AUC of 0.37407 and a test AUC of 0.55817. The model demonstrated limited precision (0.40), sensitivity (0.27), specificity (0.56), and accuracy (0.40).
ResNet50 displayed improved performance compared to EfficientNet-B3, achieving a validation AUC of 0.42099 and a test AUC of 0.58078. The model exhibited enhanced precision (0.50), sensitivity (0.30), specificity (0.67), and accuracy (0.47). Xception emerged as the most promising model, demonstrating a significantly higher validation AUC of 0.63827 and a test AUC of 0.61745. The model showcased superior precision (0.72), sensitivity (0.43), specificity (0.814), and accuracy (0.61). The results suggest that Xception outperformed the other models in accurately predicting MGMT promoter status from mpMRI data.

Table [3] presents the AUC results for EfficientNet-B3, ResNet50, and Xception, highlighting their discriminative capabilities.

| Model | Validation AUC | Test AUC |
|---|---|---|
| EfficientNet-B3 | 0.37407 | 0.55817 |
| ResNet50 | 0.42099 | 0.58078 |
| *Xception | **0.63827** | **0.61745** |

Table [3] AUC results for the three proposed Models

Table [4] provides a comprehensive overview of the performance metrics, including precision, sensitivity, specificity, and accuracy for each model.

| Model | Precision | Sensitivity | Specificity | Accuracy |
|---|---|---|---|---|
| EfficientNet-B3 | 0.40 | 0.27 | 0.56 | 0.40 |
| ResNet50 | 0.50 | 0.30 | 0.67 | 0.47 |
| *Xception | **0.72** | **0.43** | **0.814** | **0.61** |

Table [4] Performance Metrics of the three proposed Models

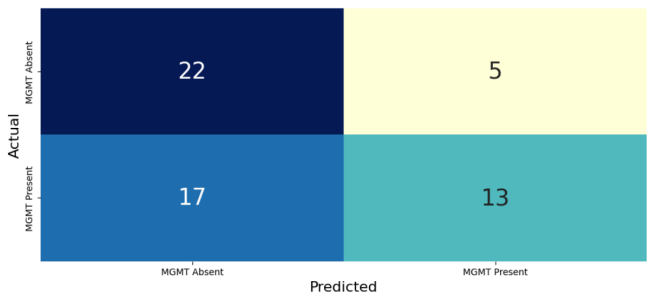

Figure [3] Confusion matrix for Xception model

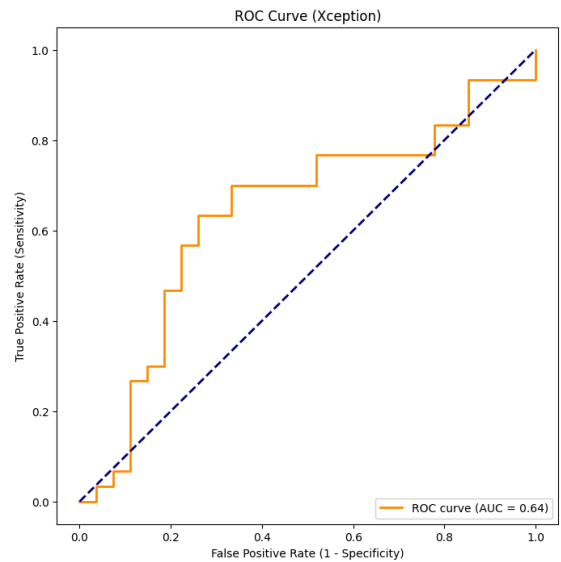

Figure [4] AUC curve for Xception model

These findings underscore the importance of exploring diverse deep learning architectures, with Xception demonstrating notable efficacy in the classification task. The diverse ensemble of models employed in this study contributes to the robustness and generalization of the proposed approach.

## Conclusion

In our exploration of predicting MGMT biomarker status in glioblastoma using diverse deep learning models, the 3D Vision Transformer (ViT3D) with a patch size of (32x32x32) and simple averaging ensemble demonstrated superior performance, achieving a testing AUC of 0.6015. Xception, among EfficientNet-B3 and ResNet50, emerged as the most promising model, showcasing a testing AUC of 0.61745 and superior precision, sensitivity, specificity, and accuracy.

These findings highlight the efficacy of state-of-the-art deep learning architectures, particularly Xception, in accurately predicting MGMT promoter status from mpMRI data. While our study contributes to the ongoing efforts in precision medicine, further enhancements can be achieved through the exploration of diverse strategies and datasets.

## Acknowledgements

The dataset for this challenge has been collected from institutions around the world as part of a decade-long project to advance the use of AI in brain tumor diagnosis and treatment, the Brain Tumor Segmentation (BraTS) challenge. Running in parallel with this challenge, a challenge addressing segmentation represents the culmination of this effort.